# From the two notions of paradigm and reduction between theories to a new multilinear History of physics

Antonino Drago
University "Federico II" of Naples - drago@unina.it

In last century, the historians of physics improved their historical accounts till up to obtain be interpretative accounts. The two main interpretative accounts have posed fundamental problems. Koyré: whether in theoretical physics mathematics is also idealistic in nature; Kuhn: whether and how the notions of paradigm, anomaly, crisis, scientific revolution and incommensurability are essential for a deep understanding of the history of physics. For their part, the philosophers of science have suggested a program for unifying the entire science; hence, they attributed to the concept of reduction between two theories insisting on the same field of phenomena a crucial role. A great debate tried to define this notion of reduction. Since longtime a particular, but more accurate notion of reduction has been applied by physicists: the reduction through a limit of a fundamental parameter of the reducing theory. But Berry, Rohrlich and Batterman pointed out that this reduction is impossible when the limit is singular, as it occurs in te cases of physical optics and geometric optics, statistical mechanics and thermodynamics, quantum mechanics and classical mechanics, etc. Hence, to represent an entire theory as the final point of a singular limit operation applies idealistic mathematics more than what was suggested by Koyré, i.e. to represent a physical law through an idealistic mathematical notion. In addition, a new mathematics – the constructive one - characterizes a singular limit as undecidable. Hence, a singular limit between two theories actually represents a difference between two different kinds of mathematics. This particular situation suggests a mathematical definition of the notion of incommensurability. As a consequence of the resulting incommensurabilities among many couples of theories the foundations of physical theories are pluralist, not only in both epistemological and ontological senses, but also in mathematical sense. Hence, the traditional vision of the historical growth of theoretical physics - as a series of theories as concentric circles, each theory being compatible with the previous ones - is denied; since longtime the history of physics is developing along a plurilinear path.

**Keywords**: New historiography of physics. Reduction between two scientific theories The debate on the definition of reduction. Limit reduction. Incommensurability. Pluralist foundations. Plurilinear history.

## 1. Introduction

In 20$^{th}$ Century on one hand historians of physics began to interpret their historical subjects by using also philosophical categories, and on another hand philosophers of physics began to formally (i.e. logically) interpret scientific theories. In the following I will compare their contributions which implicitly converge and eventually lead to a new vision of the history of physics.

In sect. 2 I will present a quick review of the various kinds of historiographies of 20th century physics, when the "new historiography of physics" was born; in particular, I will recall Alexander Koyré, Thomas S. Kuhn and Paul K. Feyerabend. In the following section I will examine how the past philosophers of physics encountered and discussed the problem of how to define a reduction between two theories insisting on the same field of phenomena. This concept is of great importance because a suitable definition of it can link all theories into a chain or a network. In sect. 4 I will provide a quick retrospective review of past proposals of reduction between two theories of classical physics; it shows that this concept can be applied to only few physical theories. In sect. 5 I will briefly summarize the debate that occurred for accurately define this concept, which however is still unclear. In sect. 6 I will examine the notion of reduction through a mathematical limit between two theories; this notion, by relying on mathematics, promised a clarification of the philosophical notion of reduction; instead, it posed further problems (the singularity of its limit), as Berry already in 1981 and Rohrlich in 1989 stressed. In sect. 7 I will present the most important case-studies of couples of physical theories commonly considered as connected by an limit reduction; almost all of limits result singular limits. In sect. 8 I will present the debate about whether the case of a singular limit makes impossible the reduction or not. In sect.9 I will link the difference between regular limit ad singular limit to the wider difference between classical mathematics and constructive mathematics. A mathematical definition of incommensurability for this case of a singular limit will result. In sect. 10 I will draw the consequences of this result for the history and the philosophy of physics; first of all the confirmation and extension of Rohrlich's pluralism in the foundations of physical theories. In sect. 11 I will conclude by remarking that the above results have been established by a great, albeit indirect, collaboration among the scholars of many different disciplines. Eventually I suggest to renew physics teaching about both the usual limit reductions which can no longer be presented as assured results and the history of physics which can no longer be presented as a unilinear path or tending to a unity.

## 2. The "new historiography of physics"

The historiography of physics began three centuries ago as a simple chronology of both the discoveries by the most important scientists and their lives. It continued by accumulating the historical re-constructions of all significant events of past theories. Eventually, in the 1930s, a "new historiography" of physics was born; it interpreted history of physics through also philosophical categories. In the year 1939 it was inaugurated by Koyré's book *Etudes Galiléennes*; then in 1969 Kuhn's book *The Structure of Scientific Revolutions* gave academic relevance to this field of study.

The former historian interpreted the birth of modern science occurred in the $15^{th}$-$17^{th}$ Centuries by taking in account as fundamental aspects of his history the cultural conception of the world of the time under examination (in particular the ideas of the dominant culture) and the relationship between physics and mathematics, which played a crucial role for the birth of physical theory. The latter historian made use of some attractive categories - normal science, paradigm, anomalies, incommensurability, crisis, scientific revolution, Gestalt - for vividly summarizing the entire history of classical physics through its salient points. But later the study-case of the beginning of quantum mechanics - black body theoretical revolution - proved irreducible to Kuhn's new categories.(Kuhn 1978) So that the new historiography has left as unexplained the births of special relativity and quantum mechanics. Since physicists did not receive any explanation of the traumatic transition of theoretical physics from classical to modern times, history of physics has then lost attractiveness and relevance.

However, the new historiography has the great merit of having posed fundamental problems; they can be summarized as follows.

1) Koyré stated that at the birth of modern time mathematics played also a platonic, idealistic role. Most scholars opposed this thesis because, being based on strictly experimental laws, theoretical physics cannot appeal to idealistic mathematical notions.

2) Kuhn's interpretation of the whole history of classical physics has proved very suggestive; yet, what do his categories mean in a specific period of time? In particular, what both the concept of paradigm and that (also suggested by Feyerabend) of the incommensurability of two theories exactly mean?

## 3. Philosophers of physics: the reduction between theories

In 1902 David Hilbert launched a grandiose program which seemed to going to solve the millennial problem of the foundations of science through the axiomatization of all theories, including the physical ones - problem 6 of his famous 23 problems to be resolved (Hilbert 1900). Hilbert's fame - due to his very important contributions to the advancements of the 20th century science - generated a general expectation of quick resolutions of all the fundamental problems of the above mentioned list, including the problem of axiomatizing all physical theories. In anticipation of the success of this grandiose program, a reduction between two of theories seemed to consist in a simply deductive relationship between their sets of axioms.

Yet, some scholars have later recognized that an experimental theory cannot be completely axiomatized, because no axiom provides the connections between theoretical assumptions and experimental data, since the latter ones cannot clearly be formalized.[1] Furthermore, in 1931

---

[1] The physicist-philosopher Mario Bunge (1972) has persisted in proposing an axiomatic approach more than others. In current mathematics, mathematical logicians apply a notion of reduction between mathematical theories

Goedel's theorems denied that also that Hilbert's program of axiomatize mathematical theories may be successful. But most mathematicians, believing there was no alternative, did not dismissed Hilbert's axiomatization, which was mantained as the best philosophical trend.

Furthermore, at the beginning of the last century a new movement in the philosophy of science seemed to revolutionize the conception of a physical theory. The logical neo-positivists proposed to reconstruct a scientific theory on the basis of its experimental data as interpreted by means of only mathematical logic (the classical one, of course). Representing the same method of understanding a physical reality, two scientific theories should necessarily be reducible to each other. Notice the following statement from the leader of this group of scholars:

> […] science is a unity, [such] that all empirical statements can be expressed in a single language, all states of affairs are of one kind and are known by the same method. (Carnap 1934, p. 32)

Assuming this program will be implemented shortly, they have launched the project of achieving a "Unified Science". But, shortly after launching this program neo-positivists admitted that they had underestimated the difficulties to be overcome. Subsequently, none suggested a reformulation of this program; which however remained as a trend for most scientists, who as their first step proposed to apply reduction between as most as possible theories. This reductionism has been proposed in an authoritative way by Sober:

> The following two theses [...] form at least part of what reductionism claims:
> (1) Every singular occurrence of a phenomenon that a higher level science can explain also can be explained by a lower level science.
> (2) Every law in a higher level science can be explained by laws of a lower level science.
> The "can" in these claims is supposed to mean "can in principle", not "can in practice". Science is not now complete; there is a lot that the physics of present fails tell us about societies, minds and living beings. However, a completed physics would not be this , or so reductionist asserts […]. (Sober 1999, p. 543)

In addition, a philosophical attitude was born, physicalism, which conceives the whole of science as a series of levels, of which physics is the fundamental one, because this science is the most experimental in nature, through QM and particle theory refers to the most elementary entities and is the more consolidated one. Moreover, most physicists maintain a tension towards a unity of the entire science; all scientific theories should be reduced in some way to one: "the theory of everything". A recent philosophical program of this kind is that of Oppenheim and Putnam (1958).

As a fact, most scientists have been fascinated by the reductionist perspective promising a unitary science, because it may achieve a web of theories, which in pairs are (at least partially) deductively connected.

> Thus, from the beginning, the discussion of issues of reduction was tied to *scientific unification*, to the epistemology of theory relations, and to [the common] scientific practice. (Van Riel 2014, p. 154; my emphasis).

**4. A quick historical review of the clamed reductions between physical theories**

In order to refer the debate on reduction to exemplary instances, let us consider those of classical physics. Notice that in the following I will consider only the reductions of completed theories; other cases (parts of theories, groups of laws, conceptions), which unfortunately are considered by various scholars discussing the theme of reduction, would make the analysis imprecise a priori.[2]

---

according to the proof theory approach (e.g., see Hofweber 2000). For a first introduction to the subject of theories reduction see (Batterman 1995; Batterman 2016).

[2] For example, Karen Crowther (2020) tries to review the feld of research; but the broad definition of the subject ("*fragments of theories*", p. 1439) is the main reason why she does not achieve clear conclusions; others reasons are her definition of reduction too wide in scope (she estabishes merely "correspondences" of these fragments; moreover, they "establish the in *principle derivability*", or the "*in principle computability*" (p. 1457) which constitute very loose conditions of reduction), as well as her missing the notion of reduction *via* a singular limit, which we will see that plays a crucial role in the entire debate on the subject of theory reducion. For a first introduction to the subject of theories reduction see (Batterman 1995; Batterman 2016).

In the history of physics, the first case of reduction of one theory to another was that between geometric optics and Newton's mechanics. Conceiving light rays as trajectories of material quanta, he interpreted the laws of reflection, refraction and even diffraction through the laws of mechanics; so after Newton geometric optics became a part of mechanical theory. A few decades after Newton's death, the invention of the differential equation of waves (1750) led to interpret as mechanical oscillations of bodies all phenomena concerning acoustics. So the physics of the phenomena perceived by the three main human senses - touch, sight and hearing - all had been mathematically interpreted by Newton's mechanics.[3] Moreover, the most representative force, namely gravity, was declared by him of a universal nature, that is, acting, in principle, in all kinds of phenomena, including chemical ones; hence, even before its birth, chemistry had to be reduced to mechanics. For these reasons, during the eighteenth century, physicists conceived of theoretical physics as a single physical theory, mechanics.

Yet, after the end of the 18th Century surprising theories emerged. Classical chemistry was born when Lavoisier put aside Newton's gravitational force. Furthermore, in the early 1800s the phenomena of polarization and interference denied Newton's hypothesis on the material constitution of light and therefore also the aforementioned reduction between the two corresponding theories. A few decades later thermodynamics and electromagnetism were born; both had fundamental concepts and laws essentially different from those of mechanics. The former unity of the entire theoretical physics was radically shake; but physicists persisted in considering the theory of mechanics as the universal model for theoretical physics. Indeed, in the second half of 19th century Kelvin strongly supported the need of explaining every non-mechanical phenomenon through a suitable mechanical model; and Boltzmann wanted to eliminate thermodynamics by reducing it to his statistical mechanics. Furthermore, a surprising case of theory reduction constituted a great success of Maxwell's electromagnetism: physical optics, whose special case in the case its waves were considered for the radii of their fronts is geometric optics, was in its turn reduced to special case of the electromagnetic waves.

But in 20$^{th}$ century the births of the first two modern physical theories - special relativity and QM - radically changed the basic notions with respect to those of classical mechanics (= CM) and any other theory. This fact, and later the births of many new, even more divergent theories made theoretical physics a markedly pluralist theoretical enterprise. After the births of such new theories, the unity of physics appeared to most as irremediably lost.

However, in the same decades of the birth of QM an important case of reduction was claimed by most physicists. Mendeleev's table of elements was interpreted by Bohr's theory of the atom and then by the subsequent theoretical and computational advances of QM. This reduction gave considerable importance to the "project of the Unified Science" of the logical neo-positivists of that time. Moreover, special relativity, apparently a "revolutionary" theory in comparison to CM, seemed to reduce to the latter theory by the limit of the light velocity $c \to \infty$. While Bohr's correspondence principle and mathematical calculations for the limit of Planck constant $h \to 0$ seemed to reduce QM to CM.

In retrospect, we see that in this history even cases of false reductions (those by Newton and Kelvin) occurred. Furthermore, the alleged reduction of thermodynamics to statistical mechanics has been strongly contested. All in all, the commonly considered successful reductions are few. All this suggests that in theoretical physics the notion of reduction, although it born three centuries ago, al present can refer to at most a little number of cases. As one authoritative scholar wrote:

> reductions in the physical sciences ... are rare and depend on special requirements. (Schaffner 2006, p. 378)

A scholar wrote the following appraisal about reduction within the entire science:

---

[3] Moreover, by a calculation on little particles Newton obtained also a first approximation of the isotherm law of gases.

If one looks for examples of reduction from the history of science, strictly derivational reductions are few and far between. (Sklar 1967, p. 110)[4]

Another scholar added:

the widespread enthusiasm for reduction can hardly be a mere induction from his past successes. (Fodor 1974, p. 97)

On the other hand, since the year 1972 a detailed opposition to reductionism has arisen; through a famous paper, Philip Anderson opposed to the reductionist program broken symmetries, at least in his field of many-body physics:

The constructionist hypothesis breaks down when confronted with the twin difficulties of scale and complexity. The behavior of large and complex aggregates of elementary particles, it turns out, is not to be understood in terms of a simple extrapolation of the properties of a few particles. Instead, at each level of complexity entirely new properties appear, and the understanding of the new behaviors requires research which I think is as fundamental in its nature as any other. […] At each stage entirely new laws, concepts, and generalizations are necessary, requiring inspiration and creativity to just as great a degree as in the previous one. (Anderson 1972, p. 393)

So the current situation of the reductionist program is very far from the goal of the unification of science. In this situation of uncertainty, it is therefore of crucial importance to clarify the definition of the notion of a reduction between two theories.

### 5. The philosophical debate on the definition of reduction between two theories

On the subject of a precise definition of reduction, there exist a number of papers suggesting proposals, variants and criticisms.

The idea that reality is composed of physical particles seems to lead directly to consider classical chemistry as a special case of QM. This is an example of a so-called *ontological reduction*, according to which one is content to explain a theory by reducing its components to the elementary ones of the reducing theory. Yet, the debate on the reducibility of the two aforementioned theories is still heated and does not seem to lean to an affirmative answer.(Drago 2020) A more general concept is that of *epistemological reduction*; the theories are compared through much more than the material constituents: laws, mathematical techniques, differential equations and their solutions, concepts, abstractions, etc. But this definition of reduction is too broad to be operational.

The logical neo-positivist philosophers, who hoped to describe every scientific theory with experimental data plus (classical) logic, saw as an ideal reduction the one between homogeneous theories for which all the laws of theory T' are deducible from those of theory T (by Schaffner was called "Nagel-Woodger-Quine paradigm"; Schaffner 1967, p. 138). But (apart from the completeness of the neo-positivist description of a physical theory) there are no historical cases of this kind. At the beginning of the debate on reduction scholars considered the case of the Galilean law of falling bodies, as reduced by logical and mathematical deduction from the laws of Newton's mechanics; but (apart from the fact that Galilei's law is not a whole theory) then Feyerabend pointed out that for Newton the acceleration of gravity is (slightly) variable, rather than constant, as assumed by Galilei.

Therefore it was necessary to consider two inhomogeneous theories (that is, some notions of the reduced theory T' are missing in the reducing theory T). The most famous definition of this reduction has been suggested by a logical neo-positivist, Ernst Nagel (1961, pp. 352-355). He conceived the two theories as two vocabularies of concepts and laws, which could be deducible from each other. In the case of inhomogeneity of the theories, the gap between the two theories is filled by introducing "bridge laws" between the terms of T' which are lacking in T and some notions of T; e.g. the law stating that the macroscopic electric current is a flow of microscopic electrons; or in statistical mechanics the laws for recovering the thermodynamic temperature, which can be defined through the average of the kinetic energy of the particles.

---

[4] A proponent of the anti-reductionism of chemistry at QM, Hans Primas, went further: "*There is no a single physically well-founded and nontrivial example for the reduction in the sense of* [...] Nagel". (1988, p. 83)

The main objection to Nagel's definition was Feyerabend's one (1962). On the base of the radical variations in meaning of the basic notions of two theories concerning the same field of phenomena both Feyerabend and Kuhn independently suggested the notion of incommensurability of these two theories. Feyerabend stressed that as a rule the couples of theories considered by the suggested cases of reduction are incommensurable; moreover, any adjustment of the two theories – or worse the introduction of an intermediate theory - aimed at obtaining their reducibility is bounded by the experimental nature of the theories at issue. Therefore non-idealistic reductions are no more than exceptions.

Owing to this opposition, later Michael Berry remarked:

… this problem of reduction has been studied a great deal by philosophers. Sometimes the discussion centres on the conflict between the two views summed up by the terms 'correspondence' and incommensurability'; in brief two theories correspond if one can be deduced as a special case of the other, and are incommensurate if their foundations are logically incompatible. (Berry 1994, pp. 597-598)

However, Monique Lévy objected that if one introduces the notion of incommensurability

One cannot understand how such [incommensurable] theories could be connected and compared, that means that it is not possible to give an acoount of the historical development of the theories and the scientific progress.(Lévy 1979, p. 340)

But Feyerabend had explained that incommensurability does not mean contradiction, nor untraslatability, nor non-cumulativity of science; it simply means a relation which is no more than partial and somewhat indirect. (Feyerabend 1979)[5]

But his concept of incommensurability has been considered as obscure by the philosophers of science and at present is highly contested.

By leaving apart this philosophical question, let us come back to Nagel reduction. For example, a paper (Fodor 1974) suggested the case of multi-realizations: eg to a macroscopic situation of thermodynamics many microscopic situations correspond (for example, the great number of microscopic ones corresponding to the definition of an entropy value $S = k\ln P$)[6]. In response to such difficulties the scholars have suggested many variants of Nagel's definition, either by relaxing the condition of deducibility (e.g. a merely descriptive explainability, or a homeomorphism, or an analogy) or by introducing - as first Schaffner (1967, p. 144) suggested - a suitable secondary theory between the two under consideration.[7] A number of definitions (or "models") resulted.

From the beginning, one scholar observed how great this variety of meanings of "reduction" is:

reductions are a very diverse bunch of items [because] reductions are distinctive scientific accomplishments of very different kinds. Some are homogeneous, other inhomogeneous. Some are 'total', others merely 'partial', Some are simple derivations, others elaborate identifications. Worse yet, some provide deeper confirmation for the reduced theory, while other serve to eliminate the reduced theory as a viable competitor for the status of scientific truth in the very act of reducing it to another accepted theory […] (Sklar 1967, p. 124)[8]

About the intense debate which followed Batterman remarked that:

---

[5] In the case of reduction beteen chemistry and QM Lombardi and Labarca (2005, p. 146) suggested a similar viewpoint, because they advocate a "coexistence of different, but objective theory-dependent, ontologies".

[6] Even more this phenomenon occurs in the relations between non-physical science: "at least some mental kinds can be multiply realized—that is: instantiated in different physiological kinds— [therefore] an identification of mental with physiological kinds is impossible" (van Riel and van Gulik 2019, sect. 3.2)

[7] For an update on the subject see (Sarkar 2015) and (van Riel and van Gulick 2019). A frequently used definition is Jaegwan Kim' notion of functional reduction, (Kim 1999) concerning the notion of causality; but both Batterman (2002, p. 127) andAlexander Rueger (2006) showed that it is not applicable to physical theories.

[8] The problems are greatly simplified if a Suppes-style axiomatic of a theory is defined, by means of a predicate that represents the theory through a model of set theory, applied to models, again of set theory, of the most representative experimental situations. In this artificial framework it is possible to establish a complete network of physical theories. (Balzer, Moulines, Sneed, 1986) through some reductions (by being satisfied with their approximate definition; Niebergall 2002. Recently, in order to obtain more precise (and also broad) definitions of reduction an attempt has been made to define a topological distance between theories understood in a structuralist sense; Gutschmidt 2014)

Showing how these derivations [according to the various models of reduction] are possible for "paradigm" examples of inter-regional reduction turn out to be rather difficult. (Batterman 2016, sect. 1)

The debate was also of a so abstract nature that even in the title of his paper Wimsatt invited to honesty, i.e. to make methodological reduction honest [, because] methodological reductionists practice "wannabe reductionism". They claim that one should pursue reductionism, but never [they] expose how [this task has to be performed]."( Wimsatt 2006, p. 445)

Worse, the debate is still inconclusive. Already in 1979 an author concluded a review on the subject of the definition of reduction by stating a negative result:

It is very clearly shown that any universal definition of reduction is impossible (Lévy 1979, p. 343)

After that time, the number of definitions grew up also because scholars' attention has widened to more "theories": biology, neurophysiology, cognitive sciences, theory of mind, etc., according to a physicalist conception that sees all theories arranged on various levels, of which the physical theory would be that reducing all the others. However, it is a fact that after sixty years of intense debate, at present no common definition of reduction is known, nor is there a dominant one. [9] An author concluded:

"The unitary Nagel-Schaffner account of reduction has dissolved, leaving a polyphonic disunity." (Wimsatt 2006, p. 447)

As a remark on the four above sections, notice that between philosophers of science, formally interpreting given theories, and historians of science, making use of philosophical categories for interpreting their historical study-cases, a unresolved conflict arose on the concept of incommensurability, rejected by the former ones and not excluded by the latter ones.

## 6. The reduction through a limit, or physicists' reduction: Rohrlich

A decisive novelty has been introduced by some physicists, exploiting the strong support that the mathematics gives to their theories and philosophies. They have conceived the concept of reduction between theories mainly in mathematical terms which i) include the logical derivations; ii) with respect to the logical deductions between axioms give one more chance to reduction, ie a limit process; iii) moreover are not disturbed by interferences with vague philosophical issues, so that the results are clear-cut answers. Surely not all possible reductions can be performed in this mathematical way, but a limit reduction constitute a necessary condition for the corresponding philosophical reduction.

This limit reduction was actually introduced by Newton, who, as remembered in fn. 2, started to reduce kinetic theory of gases to a special case of mechanics, i.e. that concerning N particles, with N tending to infinity. Well-known historical examples are the following ones: Physical optics (and Electromagnetism) and geometric optics; statistical mechanics and thermodynamics; special relativity and CM; Quantum mechanics and CM.

In the debate on the concept of reduction in 1973 Nickles pointed out that physicists reduce theories by passing to the limit of a basic parameter. It can be said that he introduced the **R scheme**:

$$\lim, \text{ for } \varepsilon \to 0, \text{ of } T = T\,';$$

for example, from QM to CM with the limit $h \to 0$.[10]

---

[9] Evidence for its insuccess is often considered the *emergence* of new phenomena or properties in T' with respect to T. But the subject is greatly debated. A property of a theory is said to be emergent if it is a new outcome of some properties of the other theory and their interaction, while it is itself different from them; hence, the explanation of the reducing theory is insufficient. When studying a reduction between a theory of microscopic world and a theory of macroscopic world this notion on one hand excludes dualism, rejecting the micro-dependence of some entities, and in the other hand reductionism, rejecting macro-autonomy. Butterfield proved that emergence is decoupled with reduction: "[…] we can have emergence with reduction, as well as without." (Butterfield 2011 b, p. 920). Being the main purpose of present paper an analysis the notion of reduction, the debate on emergence is omitted.

[10] Here a change of diction occurs. While philosophers speak of it in an active sense for the more general theory (this one reduces another theory to its particular case), physicists, following the idea of the mathematical limit, speak of it in a passive sense (the more general theoryis  reduced to a less general one).

On this subject the intervention of an authoritative physicist, Fritz Rohrlich played an important role. His first paper (1988) starts from a cosmological consideration. The whole world was born from a big-bang and then it specialized in myriads of different objects (particles, molecules, living organisms, etc.), which man conceived through different concepts (derived from abstractions and approximations with respect to reality); which he then linked in different theories. That is, specializations were born according to different cognitive levels, which in addition to specific objects gave particular meanings and different ontologies. In conclusion, despite all these forms and kinds of description, there is a substantial unity due to the historical evolution of everything from a few primordial elements; however this unity is only cosmological; for this original unity concerns the development of the universe; which has actually given rise to an ontological pluralism, as it appears to our knowledge of the present stage of world evolution. So, Rohrlich says no to ontological monism, which reductionists support to the bitter end; but yes to a monism in ultimate principle.

According to Rohrlich, in order to define a reduction, one must refer to entire theories, by excluding mere hypotheses, conceptions and models. In addition, it is necessary to distinguish three kinds of theories: accepted, if it substantially agrees with the experimental data: mature, if it is verified by experimental and theoretical practice, and finally stabilized, if its limits of validity are known. He excludes the first kind of theory; being unstable, it is not a valid object of study.

He also distinguishes the following elements in a theory: 1) the mathematical and logical structure, M; 2) the domain of validity, D; 3) the language, both informal and formal, L; 4) the epistemic components (concepts, etc.), E; 5) the ontological components (semantics), O. The mathematics of a physical theory, M, is very important, because it is the clearest component and can be handled with precision; but it represents only part of the theory. Therefore, although in a first time one may refer the notion of reduction to only the mathematical part of the theories, then he has to take into account all other components.

Analyzing the historical cases of alleged reduction, Rohrlich notes that this notion does not mean (except occasionally) the replacement of one theory for another. Moreover, the limit theory T' may be: ontologically different (eg chemistry with respect to QM), or more practical (eg Newtonian mechanics with respect to special relativity); or simpler (e.g. thermodynamic gas laws with respect the laws of kinetic theory of gases); or truer than the theory T (e.g. the cooperative phenomena suggested by the reduced theory may not exist in the theory based exclusively on elementary components, just as an excessively grainy photograph no longer makes a figure recognize).

As first, he studied the mathematical part of the two theories. He wonders under what conditions the entire mathematical framework M of a "more refined" theory is reducible to the mathematical framework of a "coarser" theory by passing to the limit on a given parameter.

He rightly remarks that:

1) Which is the parameter allowing the reduction is not always obvious.

2) The parameter must be dimensionless, because it makes no physical sense to put a physical *constant*, eg Planck's constant $h$, as a hypothetical parameter (of what?) that goes to zero; or the *constant* speed of light $c$ as a hypothetical parameter that goes at infinity: therefore not $h$, but $h/S$, where $S$ is an opportune action; not $c$, but $v^2/c^2$.

3) It is not to be hoped that the reduction can be represented by a mathematical function, nor by a functional, because physical theories are too complex with respect to a simple set of mathematical variables (or functions).

4) In order to maintain the coherence of the theory it is necessary that the energy is limited, because when eg. a particle approaches the source of a field, gravitational or electric, the potential $1/r$ goes to infinity and this final value has no physical sense.

Moreover, by distinguishing "cognitive levels" of a physical theory, Rohrlich's recognized "qualitative differences" - ie radical variations - in the meanings of the basic notions shared by the two theories. He attributed them to the "ontologies" of the two theories at issue.

By distinguishing the ontological part O from the logico-mathematical part M of a theory he states:

> Is there a reduction of the ontological component O(coarse) from O(fine)? Such a reduction implies a derivation of the O(Coarse) from O(fine). I claim that this is in general not possible. Nor can one deduce O(Coarse) from M(fine) [because one has to use] an informal language.(Rohrlich 1988, p. 310).

In such a way Rohrlich achieves an important result:

> One concludes that a [Nagelian] *logical* reduction of L(coarse) to L(fine) is not possible! Whether [for this reason] one wishes to deny [in toto] the reducibility for this reason [as Feyerabend does] or accept the mathematical reduction [at least] of M as sufficient reason for calling a theory 'reducible', now becomes a matter of definition [of reduction]. (Rohrlich 1988, p. 308)

Hence, he recognizes his convergence with Feyerabend on the notion of incommensurability, but he thought to have avoided it.

> In this way one comes close to Feyerabend theoretical pluralism and at the same time one ensures a well-defined logical-mathematical linkage. (Rohrlich 1988, p. 303) [bccause] The 'logical incompatibility proposed by some (Feyerabend 1962 and 1970) [….] is only a cognitive incompatibility. The *logical coherence* between the fine and the coarse theories is ensured by the reducibility of the mathematical part of the coarse theory. The limit involved there ensures *logical continuity* and compatibility. (Rohrlich 1988, p. 307)

### 7. Instances of singular limits in cases of claimed reduction

In the past the reductions to the limit have been implicitly conceived in the most favorable case: that is, when the passage to the limit occurs on an analytic function F (a function which can be developed in a Taylor series):

$$F(x_o + \varepsilon) = F(x_o) + \varepsilon F'(x_o) + \varepsilon^2/2 \, F''(x_o) + ....$$

But the more serious problem encountered in this study on theories reduction is that in most cases the relationship between two theories is represented by a singular limit, whose final value either does not exist or it is not close to the ever more approximating values. In the 1980s, Berry (see the 1994 paper summarizing his previous works) had highlighted reductions for which the scheme R is not valid.[11]

Subsequent studies discovered that there exist many cases of singular limit between theories (in the following we will analyze the most salient ones). It is very interesting that often these singularities correspond to new physical phenomena that bring new physical knowledge and new mathematical techniques; their divergent series carry information that has physical significance. The title of (Rohrlich 1990) announces new "good physics". Hence a singular limit between two physical theories means even physical divergence marked by a field of physical phenomena.

### *7.1 Wave optics and geometric optics*

According to Rohrlich this is a paradigmatic case of singular limits.

Geometric optics of light rays may seem to be the limit of the physical optics of waves when $\lambda$, the wavelength, becomes negligible; but this reduction is prevented by a singularity. In fact for every interval around values of $x$ and $t$ the function $\psi(x, t) = \cos(2\pi(x-vt)/\lambda)$ for $\lambda \to 0$ oscillates infinitely between $-1$ and $+1$, without achieving a limit value.

S frequent study-case is the simple phenomenon of two superposing light rays; its formula is again composed by trigonometric functions whose arguments include $1/\lambda$. Therefore they generate

---

[11] In such cases the asymptotic method of renormalization "provides a kind of limiting relationship between theories at different scales despite the fact that the reductive **Scheme R** typically fails because of divergences related to singular limits. The physics at one scale is relatively independent of that at some higher energy (shorter length). In effect, renormalization is a mathematical scheme for characterizing how the structure of interactions changes with changing scale: it turns out that the domain characterized by some lower energy (or larger length) scale is surprisingly and remarkably decoupled from that of higher energies (or smaller lengths). In other words, the decoupling entails that the higher energy regime does not much effect the behaviors and character of the lower energy regimes."(Batterman 2019, sect. 3; see also Batterman 2002, sect. 4.1)

the same singularity. It may seem like a remedy to consider energy and calculate its time average; we get $2\cos^2(2\pi (x) / \lambda)$; which however is still singular at the limit for $\lambda \to 0$. By further mediating on space one eventually obtains a finite representation of the simple geometric optics phenomenon. (Berry 1994, p. 601; Batterman 2002, pp. 80-81).

In correspondence to a singular limit, new physical phenomena emerge: caustics, where energy becomes infinite; they are important to explain eg rainbow phenomena. (Furthermore, also caustics have singularities, which can be studied with catastrophe theory). (Berry 1994, p. 602; Berry 2002, p. 11; Batterman 2002, p. 81-94).

### *7.2 Statistical Mechanics and Thermodynamics*

In textbooks it is written that statistical mechanics is reduced to thermodynamics at the limit of infinite particles: $N \to \infty$. But this limit does not exist near the critical point ($P_c, V_c, T_c$), ie when liquid and vapor emerge; there the compressibility, $\boldsymbol{K} = 1 / (- V (\partial P / \partial V)_T)$, goes to infinity because the statistical fluctuations are no longer small, but occur on a large scale, so as to generate "critical phenomena". It is remarkable that in this case fractal structures emerge at all scales: the critical state itself is a fractal. Here there can be no continuity, however far we are from the particles or they are numerous. (Batterman 2005)[12]

### *7.3 Restricted relativity and CM*

Usually special relativity is linked too CM by taking the function momentum $p = mv\beta$ (where $\beta$ is the Lorentz factor) and then developing it in Taylor series. Here the function is analytic and therefore the limit is not singular. Thus there is a reduction concerning a large part of the mathematical framework M of the two theories.

But also this reduction is a partial one. Irst, it does not concerns the entire special relativity; in the limit $E = mc^2$ disappears, because in CM there is not a relationship between mass and energy. Moreover this reduction does not obtain the entire classical physics because thermodynamics is a difficult subject; furthermore it is not true for non-massive particles, eg. photons; in fact electromagnetism does not have a Galilean invariance. Nor the limit restores the entire Newtonian mechanics; are lacking the non conservative forces and (as already remarked by Goldstein 1980, p. 332) the at distance action. Furthermore, the symmetries of special relativity change because the space-time symmetry is lost. For example, light cone of Minkowsky space, widened up to the base, does not differ from the classical space, in which $c = \infty$; but this situation is not achieved by the limit of $v^2 / c^2 \to 0$. (Rohrlich 1990, p. 1401).

Not to say, the limit operation can radically change the meanings of central concepts of the theory, first of all those of space and time since the linkage space-time disappears. In addition, also the semantics O changes: the "private" time of the observer in special relativity becomes the common time of CM.

### *7.4 CM and quantum mechanics*

Textbooks write that the two theories reduce for $h$ (or better $h / S$) $\to 0$. But the limit is singular! If nothing else, because QM is a wave mechanics (albeit of probability) and therefore the same singularity of the reduction of electromagnetism to geometric optics applies.

This singularity can also be recognized through easy examples. In the case of particles encountering a potential barrier, quantum mechanics always indicates a partial, although minimal, overcoming of the barrier: instead in CM the particles overcome or not the threshold in a sharp way. Moreover, we can consider again previous case of the superposition of two equal beams, now of particles. Classically the intensity doubles; instead in quantum mechanics there is interference, with fringes proportional to $h$; the energy depends on $\cos^2\psi$, where the argument of function $\psi$ includes $1/\lambda$, which when approaching the limit of $\lambda \to 0$ oscillates infinite times. So the limit to CM does not exist, as seen in the above; it can be recovered by taking two averages, in time and space.

---

[12] Furthermore, the notion of temperature, which is not statistical in nature. (Batterman 2002, p. 64) A more detailed analysis of this alleged reduction is found in (Sklar 2015, sect. 6). Furthermore, it must be taken into account that the very foundations of statistical mechanics are unsatisfactory; for a critical analysis see (Uffink 2004).

In support of the reduction they assert, the handbooks present Ehrenfest theorem. It finds the classical Hamiltonian equations by averaging the quantum ones. But in fact that theorem requires the hypothesis that the average values of the functions are replaceable with the functions of the average values; which only applies to small fluctuations. (Rohrlich 1990, p. 1408)

In fact, between QM and CM, semiclassical quantum mechanics comes into play for explaining quantum phenomena that are close to the classical limit. (Batterman 1995) This new field of phenomena include the following cases: 1) the base state of the helium atom; 2) the spectrum of highly excited levels of atoms in strong magnetic fields; 3) chaos. (Bokulich 2008)[13]

### *7.5 General relativity and CM*

In this case the parameter on which to operate the limit is not apparent. Any $p = kG$ (with $G$ gravitational coupling constant) at limit to 0 gives a theory without gravitational sources. But precisely because of the space-mass-energy link motivating general relativity, every limitation of energy gives a limitation on the coordinates. So $p$ it is not universal in the space. Owing to this difficulty one chooses to make the components of the deviation tensor $h_{\mu\nu}$ small as long as both the gravitational potential (with respect to the energies at rest) and the kinetic energy of the point (with respect to the its energy at rest) are small; of course, a new parameter must be defined for the latter energy: $p' = v / c \ll 1$. But one cannot avoid the question: since general relativity is also the theory of rotations and special relativity does exclude rotations, can the former one be reduced to the limit of zero rotation to the latter one? If one approximates gravity to 0, $p$ gives an equation of motion for a particle with a force $\neq 0$, while $p'$ a null force. This inconsistency makes it impossible to reduce general relativity to special relativity and indicates that $p$ and $p'$ are not independent parameters. (Rohrlich 1990, pp. 1402-1404)

### *7.6 Other problems*

1) Rohrlich poses the problem of quantum measurement. It involves reducing the interaction of two quantum systems to that of a quantum system with a macroscopic one. This reduction

> is not completely understood today. [The consequent] … "collapse of the wave function" *is an unsolved problem of theory of reduction*. (Rohrlich 1990, p. 1409)

2) There can be two limits and they do not commute. Batterman (2005) studies the case of a magnet whose temperature $T$ goes to the limit $T_c$ from below; this limit does not commute with the usual thermodynamics limit $1/N \to 0$. Moreover, if in QM the lim $t \to \infty$ for the cases of chaos is not exchangeable with the limit of $h / S \to 0$. Some confined quantum systems can simulate systems in chaotic classical evolution; but not over the long term. This phenomenon occurs for example in "quantum billiards", that is, particles confined in a plane rectangle, with wave functions corresponding to classic chaotic trajectories; there exist energy-intensive regions, which are centered on classical but unstable periodic orbits.(Berry 1981).

### 8. Which lessons from the reduction through a singular limit?

First Berry, then Rohrlich and Batterman have shown that many claimed reduction relations between two physical theories have singular limits. By taking into account this phenomenon Berry concluded a paper by stating:

> Even in what philosophers might regard as the simplest reductions, between different areas within physics, the detailed working out of how one theory can contain another [by also taking into account

---

[13] Alisa Bokulich (2008) dedicated a book to the reduction of QM to CM. On the relationships between theories she proposes an "interstructuralist" attitude, which generalizes Paul Dirac's stating the entire relationship between CM and QM through analogy between commutators and Poisson brackets. But a mere analogy greatly relaxes Nagel's deducibility requirement between the two theories in question.

limit relations] has been achieved *in only few cases* and involves sophisticated ideas on the forefront of physics and mathematics today. This is because in all nontrivial reduction the encompassing theory is a singular perturbation […] of the less general one. […] and the non analyticity describe emergent phenomena in the borderland between theories.(Berry 1994, p. 605; my emphasis)

These phenomena deny a direct reduction. Some scholars suggested a definition of reduction *via* intermediate "theories". Yet Batterman concluded negatively about the application of the various conceptions of reduction of two theories related by a singular limit: in his book he offers some examples [indicating that] the neo-Nagelian reductive relations are woefully inadequate (Batterman 2002, p. 111).

Some authors disputed Batterman's conclusions. Jeremy Butterfield (2011) analyzed four study-cases in which he concluded differently from Batterman. But the first three cases do not refer to entire theories; moreover, they concern mainly the relation between singular limit and emergence. Only the fourth study-case concerns two entire physical theories, i.e. statistical mechanics and thermodynamics. There Butterfield holds the view that the use of infinity through limits does not give particular concern because their results would be justified by "pragmatic considerations":

[…] there is a weaker, yet still vivid, novel and robust behavior that occurs *before* we get to the limit, i.e. for finite *N*. And it is this weaker behavior which is physically real.(Butterfield 2011, p. 1007)

In other terms, the use of continuous limits represents a simple preference due to the convenience of using infinity in the mathematics of physics. [14]

In such a way the question is moved to a philosophical one, i.e. the use of mathematical idealizations in theoretical physics. Instead in the following sect. 9 the question will be examined in mathematical terms, after having introduced a new kind of mathematics.

## 9. From the dichotomy on the two kinds of limit to the dichotomy on two kinds of mathematics

### *9.1 Theoretical physics and idealizations*

By studying how the first laws of theoretical physics born, Koyré claimed that mathematics played an idealistic role in theoretical physics.

Already the scientists of that time debated this question. For instance, in the following case Galilei (1638) rejected the idealistic conclusions of his argumentations on experimental data. He considered a double incline based on its vertex and a little ball descending without friction from an height *h* of the left incline; the ball will achieve the same height *h* of departure on the right hand incline, whatsoever is the angle of the latter incline and hence however the finite path of the ball is along. Yet, in the case the right incline is rotated till up to the horizontal plane the ball - in order to achieve again the height *h* from which it has been released -, has to go at infinity. Galilei excluded this idealization: he considered it as a non physical situation, although ever more approximated by all the situations of the right incline forming an ever tiner angle with the horizontal line. In modern terms, Galilei rejected the final point of a limit which is singular, because never the ball on the horizontal plane will rise up to the height *h*.

---

[14] Batterman responded to Butterfield by examining the question through the renormalization group. However he added: "It is fair to say, however, that being able to understand such inter-theoretic relations via homogeneization and renormalization techniques does not entail the existence of reductive relations between the theories either in philosophers' or physicists' sense of the terms."(Batterman 2016, end of sect. 3) For this reason present paper does not deal with renormalization. One more criticism came from an analysis of not a completed theory, but the mathematical behavior of van der Pol linear oscillator. Andrew Wayne concluded that: "The presence of a singular limit is less relevant to the argument of inter-theoretical relations than Batterman and Rueger argued", because in his study-case it is sufficient to take into account "the particular details of the explanatory resources brought bear after breakdown at the singular limit, including initial conditions, boundary conditions and empirical premises in the asymptotic methods used at singularities." (Wayne 2012, p. 354) However, he did not ventured a general criterium on singular limit reduction.

In the debate on limit reduction John Norton (2012) analyzed the distinction between approximation and idealization which is suggested by considering the final result of a limit process.

As an example, he suggested a mass falling in a weakly resisting medium (where friction function is supposed as an analytic one). At the limit of null friction the situation changes in a radical way because refers to the vacuum. In general, he stressed that the above distinction is problematic.

> There are many traps in these limits - more, I assert, than the literature has acknowledged. My concern is not the widely recognized fact that the limit may be singular […] I am concerned with far more serious oddities. The limit system may prove to have properties radically different from the finite systems, violating both determinism and energy conservation. (Norton 2012, p. 208)

About this subject Wayne remarked a defect of current philosophy of science:

> The idea that highly idealized model may underwrite bona fide scientific explanations goes against orthodox views of scientific explanation. Physicists seem to appeal to idealizations in their explanations, but we lack an account that makes sense of this practice - or even a clear view about whether these idealized accounts are intended to be part of bona fide explanations. What is needed is a well-developed theory of scientific explanation that requires neither the truth of the theories being appealed to nor deductive derivation in the explanation.[…] Clearly, further work on the role of idealization in scientific explanation is needed.(Wayne 2012, p. 544)

### *9.2 The constructive mathematics*

Rohrlich, Batterman, Rueger, Butterfield, etc. all studied the problem of inter-theoretic singular limits within classical mathematics, which was by them considered the only one possible in theoretical physics. But since the years '60s a rigorous mathematics has born that excludes the idealistic mathematics, more precisely the use of actual infinity (eg the final points of a straight line, Zermelo's axiom, etc.); it rather makes use of potential infinity, i.e. the infinity obtained by the use of only finite algorithms. The new formalized mathematics is called constructive, because it constructs every object it speaks of. It declares undecidable those mathematical objects which cannot be constructed, as of course the idealistic notions are; eg a point without extension, a perfectly parallel line to another one (i.e. also at its two points at infinity), etc.

Constructive mathematics accurately recognizes many cases of idealization. In particular, it revives Du Bois-Raymond's criticism to the notion of Cauchy-Weierstrass-Dedekind's $\varepsilon$–$\delta$ limit. According to Cantor, between two points of a however small interval there exist as many points as inside a long segment or inside an entire line; therefore not a prosecution of the approximation process but only a "miracle" can lead the two points of the interval of each $\varepsilon$–$\delta$ approximation to coincide into only one point, the supposed limit point. (Kogbetlianz 1968, pp. xxiii-xxviii) On its part, constructive mathematics defines a limit only as an approximation process which does not necessarily achieve a final result (except under additional conditions).

It is remarkable that a theorem of this new mathematics characterizes in general a singular limit as a undecidable problem.

> Let a $a_n$ be a sequence with limit $l$. Suppose $R(x)$ is a proposition concerning the number $x$. If $R(l)$ is true while $R(a_n)$ is not, then there is no effective method of determining whether or not $R(x)$ is true for any $x$ (Aberth 1980, Theorem 6.1)

Therefore to determine the final situation of a singular limit process is a undecidable problem.

Batterman (2002, sect. 4.1) and many others tried to overcome the mathematical difficulty of a singular limit by applying the techniques of renormalization group. But these techniques

> rely on the existence of non-trivial fixed points which are points in a space of Hamiltonians at which different renormalization trajectories arrive after repeated iterations of a renormalization group transformation.(Palacios 2018, p. 530)

It is well-known that in constructive mathematics the theorem of the fixed-point is undecidable.(Hendtlass 2012) Hence, the renormalization group circumvents the idealization of a singular limit through another idealization.

*9.3 Constructive mathematics and physical theories*

In the above case of the double incline Galilei was right according to constructive mathematics to reject the situation of a ball infinitely running on the horizontal plane.

Instead, in the subsequent centuries theoretical physicists made use of idealistic mathematics in defining functions (eg Dirac's $\delta$ function) and in assuming axioms (eg Newton's principle of inertia which requires $F = 0$ exactly, i.e. beyond all measurement's approximations) and also, as seen in the above, in accepting the idealized results of singular limits.

However, there exist several physical theories - Lazare Carnot's mechanics, Sadi Carnot-Kelvin-Clausius thermodynamics, first Einstein's quantum theory, Heisenberg's first formulation of quantum mechanics – that have been built without making use of idealistic notions, i.e. these authors only made use (of course, unwarily) of constructive mathematics relying on only potential infinity. Therefore, a theoretical physicist founding a new theory can use one of two kinds of mathematics, with or without actual infinity, respectively classical or constructive mathematics.(Drago 1987; Drago 1986; Drago 2012)[15]

*9.4 Constructive mathematics and inter-theoretic reduction*

As already reported, in a singular limit Butterfield (2011b) distinguished the process of limit approximation and the "final" result. This is exactly the distinction introduced by constructive mathematics; which moreover can accurately decide the singularity of a limit process through theabove quoted Aberth's theorem 6.1. By ignoring this point the literature on reduction promoted Butterfied's suggestion to a "principle" (Landsman 2013, p. 383). Correctly Patricia Palacios remarked that this promotion requires a proof of its sufficiency; that she shows is not the case.(Palacios 2018, p. 532) In addition, she showed that in the case of the claimed reduction between statistical mechanics and thermodynamics, one more limit is at stake, the temporal one for achieving the equilibrium. (Palacios 2018, p. 526) However, it has to be bounded to a "reasonable" interval of time for closely approaching it. Being not accurately defined, this "reasonable value" of time is undecidable in constructive mathematics.[16]

In sum, in the aim at improving the study of singular limits between two physical theories theoretical physicists went to re-discover the limitations foreseeable by constructive mathematics.

## 10. Philosophical implications concerning the reduction and the history of physics

It may be surprising that we deal with issues which may be considered purely philosophical in nature. Yet, they are direct consequences of very specific physical and mathematical results. Exactly this lesson is drawn by Berry from the study of limit reductions whose limits are singular:

> Singular limits carry a clear message, which philosophers are beginning to hear. The physics of singular limits is the natural philosophy of renormalization and divergent series. Perhaps they [the philosophers] are recognizing that some problems of theory reduction can themselves be reduced to tricky questions in mathematical asymptotic – an extension of the traditional philosophical method, of argumentation based on words. Usually we think of "applications" of science going from the more general to the more specific - physics to widgets - but *this is an application that goes the other way: from physics to philosophy*. (Berry 2002, p. 10; my italics)

*10.1 Koyré was right*

---

[15] The above dichotomy between the two sets of theories relying on two different kinds of mathematics has been approximated Batterman (2011; 2018) who opposed the physical theories based on continuum (we can recognize in them those traditionally linked to the mathematics of actual infinity) to those based on the microscopic discrete ( i.e. the theories which are based on a discrete mathematics - eg chemistry - or which introduce it - eg Einstein's first theory of quanta).

[16] She also proved that even in the case one can recover, by asking for a "reasonable" temporal value, the limit behavior for finite *t* "one cannot recover this behavior for realistic time scale."

In the above it was proved that the mathematical link between the two physical theories established by a singular limit may give only partial connections. But textbooks presents in an idealistic way the connection of a singular limit between two theories; they attribute the entire theory to the final situation of this kind of limit (eg CM as obtained from QM for $h \to 0$). From what has been illustrated so far it is clear that this attribution constitutes a physical idealization based on idealistic mathematics.

This result answers the question originated by Koyré's claim. Since theoretical physicists make idealistic use of idealistic mathematics when attribute an entire theory to the final situation of a singular limit between two theories, Koyré's thesis concerning a weaker idealization about single laws - as in the above Norton's example of the fall of a body under weak friction -, is proved and his historical accounts are much more profound than they appeared in the past.(Drago 2017)

For instance, at the end of a paper concerning the fall of bodies and inertia principle, Koyré concluded about the idealizations within Galilei's, Descartes' and Newton's works:

> La physique de Galilée explique ce qui est par ce qui n'est pas [c'est-à-dire ce qui est approximable, mais qui peut n'avoir le point final]… Descartes et Newton vont plus loin. leur physiques expliquent ce qui est par ce qui ne peut pas être ; elles expliquent le réel par l'impossible. Galilée… ne le fait pas. [Son] impossibilité de ce mouvement [idéal]… n'a pas la même structure.(Koyré 1986, p. 276)

It is remarkable that by means of an intuitive analysis of the texts of that time Koyré recognized a radical difference between two different ways of approximating a result (we can say through a regular limit or a singular limit) and moreover he was capable to express their distinction by means of adequate words.

### *10.2 A mathematical definition of incommensurability*

About the question of which philosophical conclusion one can derive from the above analysis on singular limit reductions it is interesting to consider Rohrlich's progressive steps represented by his three papers on the subject.[17] As seen in the above sect. 5, Rohrlich recognized the variations in meaning occurring between the two theories examined under the possibility of a reduction.(Rohrlich 1988, p. 303) He knew that according to (both Kuhn and) Feyerabend this phenomenon implies the incommensurability of the two theories. Rohrlich exorcised this notion since he thought that an incommensurability implies "contradictions […] and the impossibility of even weak theory reduction"; and at last "the lack of cumulativity of knowledge". Rather, he concluded that the process of reduction - considered as a deductive process -, is impossible only between the epistemological E and ontological O parts of the two theories.(Rohrlich 1988, p. 303)

In a next paper he dealt with the difficult study-case of the limit reduction of general relativity to both special relativity and CM. He had to take into account that the relation between the mathematical parts M of the theories may include a singular limit. He concluded:

> There is of course a lot more to a physical theory than its mathematical structure; there is its language, its ontology, and its epistemology. All these components are part of the interpretation of the theory. And this interpretation of T' is in turn strongly dependent on the mathematics of T. […] One concludes that the logic of theory reduction is the logic of the reduction of [only] its mathematical structure. [Because] Mathematics controls the interpretation, the "paradigm change" in reduction, just as it controls that in scientific revolutions of physical theories. (Rohrlich 1989, p. 1169)

Here it is apparent that Rohrlich hoped that one can overcome the insufficient clarification of the ontological and epistemological differences between the two theories through the logico-mathematical part M of a theory. But he did not prove it.

One year later a third paper - devoted to the study of the "good physics" corresponding to a singular limits - presents the opposite opinion; now mathematics does not control the ontological interpretation and his conclusion about reductionist work is discouraging:

---

[17] For a first analysis of the evolution of Rohrlich's ideas see (Batterman 1995, p. 175)

I have shown elsewhere [ie previous two papers] that the semantic component of theory reduction (its interpretation) does not carry through in many cases, i.e. , that the physical model of theory T' does not follow from that of T but involves an interpretation of the mathematics of T' which is independent of the interpretation of T. If the mathematical component of theory reduction is now also found inadequate (at least for certain kinds of physical systems), there is little left of the generality of reductionism. (Rohrlich 1990, pp. 1410-1411; I changed the denotation S into T')

In the beginning of sect. 8 the similar negative appraisals by both Berry and Batterman have been quoted. Yet, these scholars did not derive any philosophical consequence. Only the first paper of Rohrlich had mentioned the philosophical question of the incommensurability, but for avoiding it.

Instead, from the application of constructive mathematics one can derive a formal definition of incommensurability. The case of a singular limit between two theories constitutes hard evidence of an incommensurability of the entire mathematical frameworks M of the two theories. A undecidable problem severs them. Moreover, a singular limit between two theories represents the most important radical variation in the meaning than the variations of all other notions or properties because it includes variations even in the mathematical formulas representing a same phenomenon – eg the energy formula of two superposing rays in electromagnetism and geometrical optics is different in the two theories. Hence, in this particular case of a singular limit between two theories we have an accurate – owing to its mathematical nature - definition of the intuitive concept of incommensurability which was suggested by both Kuhn and Feyerabend in philosophical terms. Through the case of a singular limit between two theories the concept of incommensurability is therefore rooted inside theoretical physics in a well-defined way.

### 10.3 The rarity of reductions of entire theories

From the early debate on reduction we learnt that does not exist a case of reduction-deduction between two homogeneous theories (i.e. a deduction from some laws of one physical theory to *all* laws of the other theory). Moreover, in sect. 7 we have examined at least six important cases of theory reduction through a limit; five cases rely on are singular limits whose "final" points in classical mathematics either do not exist or represent idealizations. Only one case is based on a regular limit, i.e. the reduction between special relativity and CM; which however does not obtain the entire latter theory. One may only speculate case by case whether by adding specific bridge laws or even intermediate theories a reduction is any way possible; but the result cannot be generalized, as the unsuccessful search for a general definition proves. All that, together with the above quoted conclusions by Berry, Rohrlich and Batterman allow to conclude that a reduction is rarely successful.

Already in the year 1988 Rohrlich suggested a pluralism of ontological nature. After having recognized the difficulties introduced by singular limits also mathematical pluralism is well-established; this pluralism appears to be the general rule and the case of a reduction an exception.

It is not then surprising that along sixty years the philosophical debate – among scholars whose majority is favorable to reductions – did not exhibit couples of physical theories (beyond the case of special relativity and CM) as instances of successful reductions (rather, at present times it appeals to more imprecise theories, ie the biological, phisiological or psychological ones, all lacking of an assured mathematical framework). Nor it is surprising that this debate was very abstract in nature, so much to lead Wimsatt to appeal for a more "honest" debate.

### 10.4 Devaluation of the program of unifying all theories

Already Ernst Mach (1896) invited to study the mathematical homeomorphisms between theories insisting on different fields of phenomena: eg. a RC electric circuit and a damped mechanical oscillator are represented in mathematics by a same differential equation. This link between theories leads to many useful considerations. But many scholars think that it is imaginative to generalize this link to the entire two theories, ie electromagnetism and mechanics. However, in the name of a same, universal scientific method neo-positivists claimed that all theories have to be unified. Subsequently, also reductionists and physicalists reiterated this program.

In the above an accurate analysis showed that a singular limit does not allow a reduction of the mathematical frameworks M of one theory to another theory in its entirety; it connects only parts of the two physical theories (eg in the case of QM to CM one obtains many but not all formulas of the latter one: the formulas for extended bodies, or those for motions with friction, etc.). In general, it is not known whether these common parts (eg semiclassical quantum mechanics) constitute autonomous theories, or open theories, or merely mathematical models.

Therefore, reductionists emphasizing the few successful cases of partial connections of the mathematical frameworks M of two theories as the starting point of the application of a programme reductions for all theories, are suggesting a programme that, after almost a century of so many unsuccessful attempts, has to be considered clearly unattainable; hence, their final result - the unity of the entire science - has to be considered as mythical.

After half a century of a specific research on the subject of the neo-positivist notion of reduction, Feyerabend's appraisal on the philosophy of reduction seems well-adequate:

> … a formal account of reduction is impossibile for general theories… Nagel's theory of reduction… [is] not in accordance with actual scientific practice and with a reasonable empiricism (Feyerabend 1962, p. 28)

This conclusioni is even more valid if the empirism in theoretical physics is associated to constructive mathematics.

Also Batterman has devalued the program of reductionists. He countered Sober's reductionist theses (quoted in sect. 3):

> I have argued that [his] theses are essentially meaningless.[…] I think that the ideal of *in principle* derivations of behavior of systems (or laws, or theories) from more "fundamental" lower-scale details (or more fundamental laws or theories) is largely mistaken. Any examination of the actual practice of scientists interested in modeling systems at different scales will reveal nothing as simple as the kind of derivations that the proponents of this ideal believe is possible. [Sober's] appeal to a completed ideal physics - the main feature that underwrites these in principle claim - is purely aspirational and speculative. We have no idea what such a physics would like look, nor we have any evidence that it exists. (Batterman 2018, p. 871)

*10.5 New interpretation of the history of Physics*

As a consequence, the traditional vision of the growth of theoretical Physics as a series of concentric circles - each new theory being compatible with the previous theories, since it represents an extension of them - fails. Already in the year 1988 Rohrlich suggested, on the basis of his long experience as a theoretical physicist, that there is a plurality of theories' foundations. Moreover, in the above it was proved that thermodynamics and statistical mechanics have to be seen as incommensurable, because, *pace* Boltzamnn, in classical mathematics the limit between them is singular and in constructive mathematics is undecidable. Hence, since thermodynamics' birth physical theories rely on incompatible foundations. the history of physics is no longer represented by a unilinear evolution, which disregards some "less important" theories as lateral developments; it is rather represented by a plurilinear evolution in correspondence to the different kinds of foundations.

This result denies Kuhn's vision of a single line of historical development constituted by consecutive paradigms (each one arising just after the previous paradigm through a Gestalt phenomenon within scientists' community). Moreover the plurilinear vision of history of physics attributes new interpretations to Kuhn's categories. In the history of science the birth of a "paradigm" occur when a theory of one line of development culturally dominates all others theories. Such a kind of paradigm occurred in the 18[th] Century only; it was the paradigm of Newtonian mechanics (to which all other physical theories have been successfully reduced as particular cases, of course according to the inaccurate notion of reduction of that times: geometrical optics, acoustics, plus the first laws of electricity and magnetism). In the history of physics a "crisis" occurs when theories of a subordinate line suggest a destabilizing "anomaly" (eg it occurred in the first decades of 19$^{th}$ century when phenomenological physics suggested the interference and polarization phenomena which denied Newton's reduction of geometric optics to CM; it occurred almost a century later when Lorentz' theory of electron suggested a new theoretical transformations which were incompatible with those of the mathematical framework of CM; or when thermodynamics suggested through black-body theory discrete quanta of energy, conflicting with the entire tradition of classical physics).[18] Hence an essentially plurilinear history of physics began two centuries ago, in the early years of 19th Century.(Drago 2013)

## 11. Conclusions

Some crucial problems bequeathed by historians and philosophers of scince of 20th Century have been solved by scholars who have bravely engaged themselves within research fields that were outside their specific expertise: 1) the (neo-positivist) philosophers who have payed accurate attention to the (relationships between) scientific theories till up to analyze their formal details; 2) the historians of physics (Koyré, Kuhn, etc.) who have made use of philosophical categories, suggesting very advanced accounts of the history of physics; 3) the physicists (Berry, Rohrlich, Batterman, etc.) who in order to investigate the reductions between physical theories have studied the mathematics of singular limits; 4) the mathematicians (Markov, Bishop, Aberth, etc.) who have clarified in formal terms the philosophical attitude of eliminating from current mathematics the actual infinity, and hence the idealizations. This interdisciplinary commitment of each group of philosophers, mathematicians, physicists and historians, and their - indirect and from afar – collaboration led to recognize a pluralism within the foundations of Physics.

Therefore, present physics teaching should change his careless way of presenting as assured limit some reductions which actually are based on singular limits; at least it has to warn that singular limits (esp. those between statistical mechanics and thermodynamics, QM and CM) give no more than partial connections between the mathematical formulas of the two theories at issue.

Moreover, pluralist foundation suggests a radically new history of physics, which is very different from that currently presented (or only hinted at) by physics teaching. As a matter of fact, this evolution does not converge towards a unity. Therefore present presentation of the historical evolution of the physical theories should be rethought and changed in a liberal attitude about a plurality of foundations.

---

[18] According to Kuhn the change between two successive paradigms is represented by a *Gestalt* phenomenon. In the most favorable interpretation, the notion of "Gestalt" alludes to a singular limit in the relationship between the two theories representing the old and the new paradigm (e.g. the singular limit between CM and quantum mechanics).